\documentclass[article,amsmath,amssymb,aps,superscriptaddress,twocolumn,prb]{revtex4-2}
\usepackage[normalem]{ulem}
\usepackage{graphicx}
\usepackage{dcolumn}
\usepackage{bm}
\usepackage{natbib}
\usepackage{lipsum}
\usepackage{amsmath}
\usepackage{physics}
\usepackage{multirow}
\usepackage{url}
\usepackage{textcomp}
\usepackage{soul}
\usepackage[hidelinks]{hyperref}
\usepackage{breakcites}
\usepackage{setspace}

\newcommand{\fluence}{\,nJcm$^{-2}$}
\newcommand{\mob}{\,cm$^2$V$^{-1}$s$^{-1}$}
\newcommand{\me}{\,$m_\mathrm{e}$}
\newcommand{\den}{\,cm$^{-3}$}
\newcommand{\wc}{$\omega_\mathrm{c}$}
\begin{document}
\title{Hot electron cooling in InSb probed by ultrafast time-resolved\\terahertz cyclotron resonance}

\author{Chelsea Q. Xia}
\affiliation{Clarendon Laboratory, Department of Physics, University of Oxford, Parks Road, OX1 3PU Oxford, United Kingdom}

\author{Maurizio Monti}
\affiliation{Department of Physics, University of Warwick, Gibbet Hill Road, CV4 7AL Coventry, United Kingdom}

\author{Jessica L. Boland}
\affiliation{Clarendon Laboratory, Department of Physics, University of Oxford, Parks Road, OX1 3PU Oxford, United Kingdom}
\affiliation{Photon Science Institute, Department of Electrical and Electronic Engineering, University of Manchester, Oxford Road, M13 9PL Manchester, United Kingdom}

\author{Laura M. Herz}
\affiliation{Clarendon Laboratory, Department of Physics, University of Oxford, Parks Road, OX1 3PU Oxford, United Kingdom}

\author{James Lloyd-Hughes}
\affiliation{Department of Physics, University of Warwick, Gibbet Hill Road, CV4 7AL Coventry, United Kingdom}

\author{Marina R. Filip}
\affiliation{Clarendon Laboratory, Department of Physics, University of Oxford, Parks Road, OX1 3PU Oxford, United Kingdom}

\author{Michael B. Johnston}
\email{michael.johnston@physics.ox.ac.uk}
\affiliation{Clarendon Laboratory, Department of Physics, University of Oxford, Parks Road, OX1 3PU Oxford, United Kingdom}

\date{\today}

\begin{abstract}
	Measuring terahertz (THz) conductivity on an ultrafast time scale  is an excellent way to observe charge-carrier dynamics in semiconductors as a function of time after photoexcitation. However, a conductivity measurement alone cannot separate the effects of charge-carrier recombination from effective mass changes as charges cool and experience different regions of the electronic band structure. Here we present a form of time-resolved magneto-THz spectroscopy which allows us to measure cyclotron effective mass on a picosecond time scale. We demonstrate this technique by observing electron cooling in the technologically-significant narrow-bandgap semiconductor indium antimonide (InSb). A significant reduction of electron effective mass from 0.032\,$m_\mathrm{e}$ to 0.017\,$m_\mathrm{e}$ is observed in the first 200\,ps after injecting hot electrons. Measurement of electron effective mass in InSb as a function of photo-injected electron density agrees well with conduction band non-parabolicity predictions from \textit{ab initio} calculations of the quasiparticle band structure.
\end{abstract}

\maketitle

\section{Introduction}
Hot charge carriers play an important role in semiconductor devices such as Gunn diodes~\cite{couch1989hot,forster2007hot}, avalanche photodiodes~\cite{capasso1985physics,abautret2015characterization} and hot carrier solar cells~\cite{ross1982efficiency,konig2010hot}. However, they can diminish the performance of other devices such as field effect transistors~\cite{klaassen1970influence}. Therefore, a deep understanding of the temporal evolution and cooling of hot carriers in semiconductors is an important area of research, which is linked closely to the development of a wide range of semiconductor devices.

Hot carriers in a semiconductor may arise by absorption of photons with energy significantly larger than the bandgap; by charges gaining kinetic energy, for example via application of an electric field; or via electrical injection from high-energy side valleys in the band structure, or a heterojunction. While electron and hole distributions are initially likely to be non-thermal (i.e. not able to be characterized by a Fermi-Dirac distribution), thermalization typically occurs on 10\,fs time scale~\cite{erskine1984femtosecond,taylor1985ultrafast,rota1993ultrafast} after which it is possible to define electron and hole temperatures, which are higher than the lattice temperature.  

Conduction band minima and valence band maxima can be approximated by a parabolic dispersion for energies very close to the maxima/minima, however this approximation becomes less valid at higher energy. Thus band non-parabolicity is particularly important for hot-carrier devices. For example, as a charge is accelerated to an energy beyond the parabolic region, it becomes effectively heavier as the band flattens from a parabolic shape, leading to a reduction in the charge mobility. A population of hot carriers therefore has a wide distribution of effective masses.  
  
Hot carrier effects are pronounced in many narrow-bandgap semiconductors. InSb is a typical direct-gap semiconductor, whose bandgap is only 0.17\,eV at room temperature and 0.235\,eV at 4\,K~\cite{littler1985temperature}. Its small electron effective mass, large electron mobility and high thermal conductivity make it an ideal candidate for high-speed electronic devices such as ultrafast transistors~\cite{transistor2004novel} and photodetectors~\cite{kimukin2004high,kuo2013high}. Its narrow bandgap is also suitable for infrared detection and thermal imaging~\cite{InSbThermo2005thermoelectric}. Furthermore, quantum-well heterostructures of InSb and AlInSb show a strong quantum Hall effect~\cite{InSbQuantum2012high}. Owing to these optoelectronic properties, extensive studies have been performed on InSb over the last few decades. For example Dresselhaus \textit{et al.} found that the effective mass of InSb at the bottom of the conduction band was 0.013\me\ by measuring cyclotron resonance at microwave frequencies~\cite{dresselhaus1955cyclotron}. The extremely small effective mass implies an ultrahigh charge mobility, which has been reported to be $7.8\times10^4$\mob\ at ambient temperature~\cite{rode1971electron}.

The non-parabolicity of the conduction band of InSb has also been studied extensively over the past half century. Early cyclotron resonance experiments at infrared frequencies revealed an increase of effective mass with magnetic field~\cite{infra_burstein1956cyclotron,infra_keyes1956infrared,infra_palik1961infrared} which agreed with Kane's \textbf{k}$\cdot$\textbf{p} perturbation theory model~\cite{kane1957band}. In Kane's model the introduction of non-parabolicity gives rise to an extra energy in the dispersion relation of the lowest conduction band, which is expressed as $E(\mathbf{k})(1+E(\mathbf{k})/E_\mathrm{g})=\hbar^2|\mathbf{k}|^2/(2m_0^*)$ where $E_\mathrm{g}$ is the bandgap energy, $\hbar\mathbf{k}$ is the crystal momentum and $m_0^*$ is the effective mass at the conduction band minimum, which is taken here as the  energy reference. Band non-parabolicity in InSb has also been observed through the observation of ``forbidden" Landau transitions~\cite{combine_mccombe1967combined,combine_mccombe1969infrared,combine_enck1969phonon} and the magnetic-field-dependence of acoustic absorption~\cite{wu1971effect}. Another way that band non-parabolicity has been probed is via microwave or infrared cyclotron absorption measurements of doped samples with different chemical potentials. This is usually achieved  by changing the extrinsic charge-carrier concentration. Since multiple samples with different doping densities are needed for this method, it is less convenient to probe the conduction band at a specific energy state and the range of measurable states are limited by the available doping densities of the samples~\cite{smith1959energy,infra_palik1961infrared,combine_mccombe1969infrared,johnson1970infrared}.

Free electron lasers (FELs) have been used previously to measure cyclotron resonance and hot-carrier cooling in undoped InSb with a picosecond time resolution. It was reported that the electron cyclotron mass of InSb  increases at late times after photoexcitation, an effect attributed to the renormalization of the bandgap energy~\cite{kono1999picosecond}. In contrast, similar measurements performed on InSb quantum wells showed a decrease in effective mass at later delay times. Moreover a significant initial cyclotron resonance broadening was observed at early times after photoexcitation, owing to initial Landau occupancy and high momentum scattering rates~\cite{khodaparast2003relaxation}.

More recently, magneto-THz spectroscopy has been realized to investigate cyclotron resonance with sub-picosecond resolution. Unlike the FEL technique~\cite{kono1999picosecond,khodaparast2003relaxation} where cyclotron resonance was investigated by changing the magnetic field at a fixed FEL wavelength, with the time-resolved magneto-THz spectroscopy technique, the time evolution of cyclotron resonance spectrum at a fixed magnetic field is recorded.
This technique was used previously  to study the behavior of magneto-excitons in GaAs quantum wells~\cite{lloyd2008terahertz,zhang2016stability} and Rashba spin-orbit splitting in germanium quantum wells at high magnetic fields~\cite{failla2015narrow,failla2016terahertz}. A mode softening effect of the polaritons was also reported on germanium and InSb quantum wells in the ultra-strong coupling regime where a high magnetic field was applied~\cite{keller2020landau}.

\begin{figure*}[t!]
	\includegraphics[width=\textwidth]{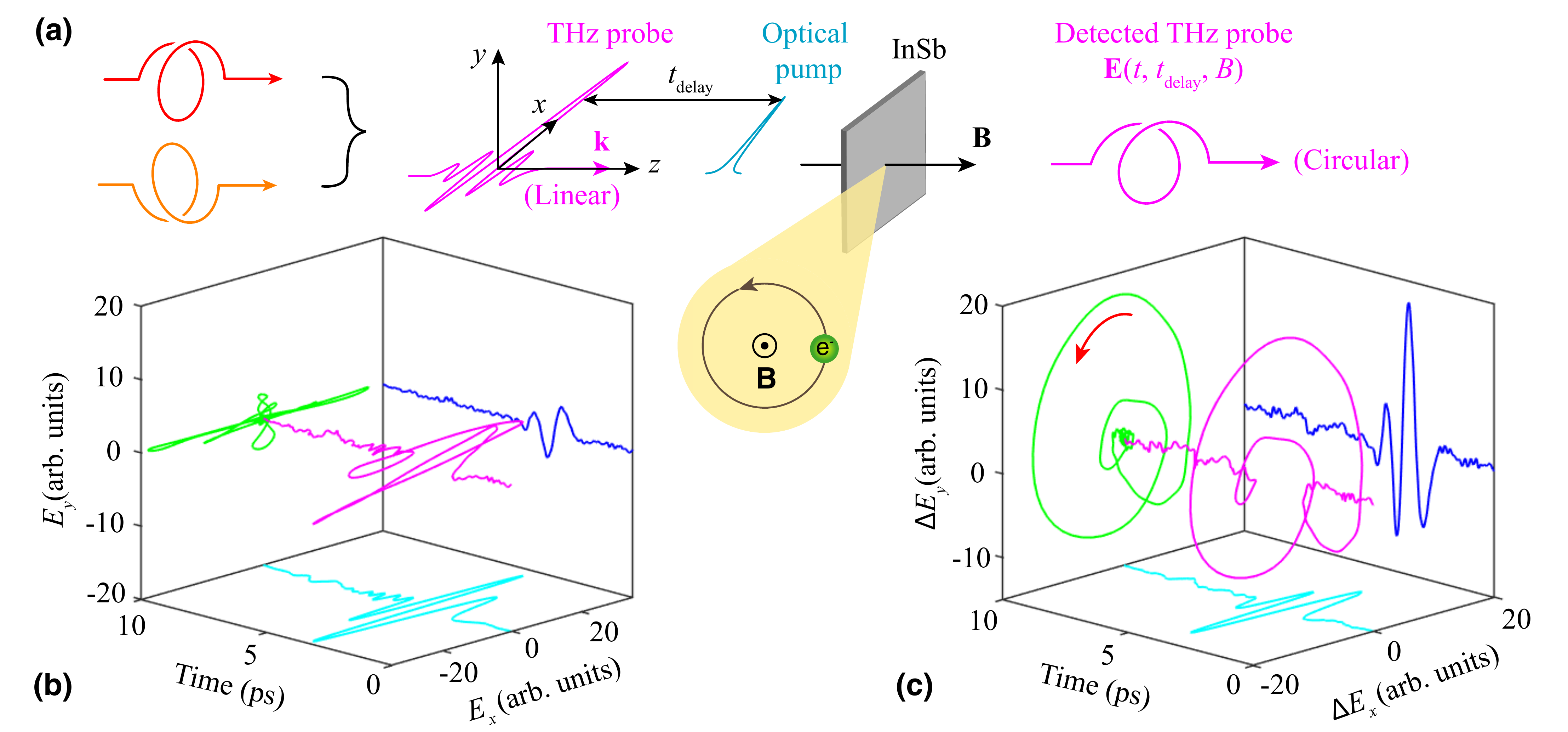}
	\caption{(a) A schematic diagram of the time-resolved magneto-THz spectroscopy experiment of an undoped InSb sample. The magnetic field vector $\textbf{B}$ and the \textbf{k}-vector of the THz pulse are in the same direction and perpendicular to the sample surface. The linearly polarized incident THz pulse is a superposition of right-handed and left-handed circular polarization components. (b) Electric field of the transmitted THz pulse without photoexcitation, $\mathbf{E}_\mathrm{dark}(t,t_\mathrm{delay},B)$, measured at a magnetic field $B=0.6$\,T. (c) The change of THz transmission $\Delta{\mathbf{E}}(t,t_\mathrm{delay},B)$ measured at 200\,ps after photoexcitation with a fluence at 5.4\fluence\ under a magnetic field $B=0.6$\,T. The $x$-component and the $y$-component of the recorded THz signal are represented by the light blue and dark blue curves respectively and the pink curve represents the resultant THz signal. The small feature seen in the THz signal at 7.5\,ps is an experimental artefact caused by a weak reflection from an optical element along the THz beam path.}
	\label{figure:cr_3d}
\end{figure*}

In this work we utilized time-resolved magneto-THz spectroscopy which is based on optical-pump-THz-probe spectroscopy (OPTPS) to observe cyclotron resonance and hot-carrier cooling in undoped InSb. Electron cooling was observed on an ultrafast timescale (up to 200\,ps after photoexcitation) by varying the time delay between an optical pump pulse and a THz probe. In addition, to gain deeper understanding of the non-parabolicity of InSb, the measured effective mass as a function of carrier density was compared with \textit{ab initio} calculations within the $G_0W_0$ approximation~\cite{hybertsen1986electron, marini2009yambo} showing very good agreement. The time-resolved THz technique allows the cyclotron effective mass, charge scattering rate and charge density to be measured with picosecond time resolution, thereby providing insight into hot-carrier cooling in InSb.

\section{Time-resolved magneto-terahertz spectroscopy}
OPTPS is an excellent non-contact probe of photo-conductivity in semiconductors. It measures photoconductivity $\Delta\sigma$ on a sub-picosecond timescale allowing the direct observation of charge cooling~\cite{bretschneider2018quantifying} and recombination dynamics~\cite{wehrenfennig2014charge,milot2015temperature}. In OPTPS, conductivity is probed using a sub-picosecond single-cycle electromagnetic (THz) pulse which travels in free space, thereby avoiding the need to make physical electrical contacts to the semiconductor. Hence the technique is well suited to nanomaterials and systems where Ohmic contacts are difficult to achieve. Time-resolved measurements are obtained by adjusting the time delay, $t_\mathrm{delay}$, between an optical pulse, which photo-injects electron-hole pairs in the semiconductor and the THz probe pulse.

In traditional OPTPS measurements a frequency averaged conductivity can be measured~\cite{wehrenfennig2014charge,milot2015temperature} at each time after photoexcitation or the full THz AC photoconductivity spectrum can be recorded at each point after illumination. In most bulk semiconductors the AC photoconductivity spectrum fits well to the Drude model,
 \begin{equation}
 	\Delta\sigma_{xx}(\omega)=\frac{ne^2}{m^*}\frac{i}{\omega+i\gamma},
 	\label{eq:drude}
 \end{equation}
 where $n$ is the charge-carrier density determined by the photoexcitation fluence, $m^*$ is the effective mass of the charge carrier
 and $\gamma=1/\tau$ is the momentum scattering rate with $\tau$ corresponding to the momentum scattering time. Thus it is possible to extract $\gamma$ and the value of $n/m^*$ as a function of time after photoexcitation. By measuring the absorption coefficient of the semiconductor and the fluence of the laser pulse it is also possible to determine $n$ and hence establish $m^*$ immediately after photoexcitation. However, at later times $n$ and $m^*$ cannot be separated as both quantities change as a result of charge recombination and band non-parabolicity respectively. Thus studying hot charge carrier relaxation with this technique is challenging. 

Here we extend OPTPS to perform time-resolved cyclotron resonance spectroscopy and hence observe $m^*$, the energy dependent cyclotron effective mass, on a picosecond timescale. Fig.~\ref{figure:cr_3d}(a) is a pictorial representation of the time-resolved magneto-THz spectroscopy experiment. A femtosecond laser pulse (optical pump) is first used to photoexcite a semiconductor sample which is held at low temperature in a magnetic field, $B$. A linearly polarized THz probe pulse with \textbf{k}-vector parallel to the magnetic field arrives at a set time, $t_\mathrm{delay}$, after the optical pump pulse and the amplitude of the electric field of the THz pulse $\mathbf{E}(t,t_\mathrm{delay})$ is recorded after it has been transmitted through the sample. Meanwhile, the electric field of the THz pulse passed through the photoexcited semiconductor is compared to that of the unilluminated semiconductor to give a change in the THz signal $\Delta \mathbf{E} (t,t_{\text{delay}})=\mathbf{E}_\mathrm{light}(t,t_{\text{delay}})-\mathbf{E}_{\text{dark}}(t,t_{\text{delay}})$, which can be measured simultaneously with $\mathbf{E} (t,t_\mathrm{delay})$. Therefore $\Delta \mathbf{E} (t,t_{\text{delay}})$ represents the change in the THz pulse associated with the photo-injection of electron-hole pairs. In the case of electron cyclotron resonance with a magnetic field in the same direction as the THz pulse \textbf{k}-vector, a right-handed circularly polarized component of the incident THz pulse will be absorbed in the process of promoting electrons to the next available Landau level. Therefore, the recorded $\Delta{\mathbf{E} }(t,t_\mathrm{delay},B)=\mathbf{E}_\mathrm{light}(t,t_\mathrm{delay},B)-\mathbf{E}_\mathrm{dark}(t,t_\mathrm{delay},B)$ directly measures cyclotron resonance of photo-injected electrons at the frequency, $\omega_{\rm c}=eB/m^*$. Both $\Delta{\mathbf{E}}(t,t_\mathrm{delay},B)$ and the cyclotron motion of the electrons are right-handed circularly polarized, as illustrated in Fig.~\ref{figure:cr_3d}(a).

An example of $\mathbf{E}_\mathrm{dark}(t,t_\mathrm{delay},B)$ and $\Delta{\mathbf{E}}(t,t_\mathrm{delay},B)$ for InSb measured at $t_\mathrm{delay}=200$\,ps  after photoexitation under a magnetic field of at $B=0.6$\,T is shown in Fig.~\ref{figure:cr_3d}(b) and \ref{figure:cr_3d}(c) respectively. By taking the Fourier transform of these two data sets, it is possible to obtain the change in photoconductivity associated with the cyclotron resonance,
\begin{equation}
	\Delta\sigma_{xx}(\omega)=-\varepsilon_0c\alpha(1+\tilde{n})\left[\frac{\Delta{E_x}(\omega,B)}{E_x^\mathrm{dark}(\omega,B)}\right],
	\label{eq:ccond}
\end{equation} 
where $\varepsilon_0$ is the permittivity of free space, $c$ is the speed of light, $\alpha$ is the absorption coefficient of InSb at the pump wavelength (800\,nm) and $\tilde{n}$ is the refractive index of InSb at THz frequencies. The resultant photoconductivity spectrum of InSb under a magnetic field features a characteristic cyclotron absorption peak at frequency \wc. This cyclotron resonance feature can also be observed as an absorption dip in the transmission spectrum,
\begin{equation}
T_x(\omega,B)=\frac{E_x^\mathrm{light}(\omega,B)}{E_x^\mathrm{dark}(\omega,B)},
\label{eq:trans}
\end{equation} 
which may be used for investigating the electron effective mass of InSb as a function of magnetic field strength as shown in Fig.~\ref{figure:cr_color}(a).

It should be noted here that given the field and polarization-resolved nature of the magneto-THz spectroscopy technique, it is possible to extract and assign electron and hole cyclotron effective masses from the same dataset according to the handedness of cyclotron resonance (if measured at a value of $B$ for which both electron and hole effective masses lie within the spectral bandwidth of the THz pulse).

For the measurements described in this paper samples were photoexcited with a 35\,fs duration pulse of 1.55\,eV photons (central wavelength 800\,nm) from an amplified Ti:sapphire laser, while the THz pulse was generated via optical rectification by shining pulses from the same laser system at a 1\,mm-thick (110) ZnTe crystal. The THz electric field was detected by a 1\,mm-thick (111) ZnTe crystal along with a half-wave plate, a quarter-wave plate, a Wollaston prism and a pair of balanced photodiodes using the method of electro-optic sampling, from which the orthogonal polarization components of the THz pulse can be separated. Further details of the THz generation and detection technique are given in Appendix A.  

\section{Cyclotron resonance in photoexcited InSb}
To investigate the band structure of intrinsic InSb,
we utilized the magneto-THz spectroscopy to measure the effective mass of electrons within the conduction band. A  6.5\,$\mu$m-thick nominally-undoped layer of InSb was grown by molecular beam epitaxy on a wafer of (100)-oriented semi-insulating GaAs. All measurements were performed with the InSb sample held at a temperature of 4\,K in a helium-flow cryostat within the core of a superconducting magnet. 
The mobility of InSb was extracted from the photoconductivity spectrum taken at 4\,K, which gave a value of $2.3\times10^4$\mob. The Hall measurements gave a carrier concentration of $1.7\times10^{16}$\den\ at room temperature, which is consistent with the intrinsic concentration of thermally excited carriers. At 77\,K, however, the obtained carrier concentration is $6.4\times10^{14}$\den\ which is much higher than what is expected from thermal excitation and indicates a background doping density of approximately $10^{15}$\den. Nonetheless, since our cyclotron resonance measurements were performed at 4\,K, we assumed that the extrinsic carriers arising from defects and impurities were frozen out and also that the concentration of intrinsic carriers is negligible~\cite{ashcroft1976solid}. Hence, we consider the InSb sample as undoped at 4\,K and the Burstein-Moss shift negligible, which gives a bandgap energy of 0.235\,eV. As a result, each absorbed 1.55\,eV photon from the laser pulse generates extremely hot electron-hole pairs with 1.32\,eV kinetic energy.
\begin{figure}[ht!]
	\includegraphics[width=0.5\textwidth]{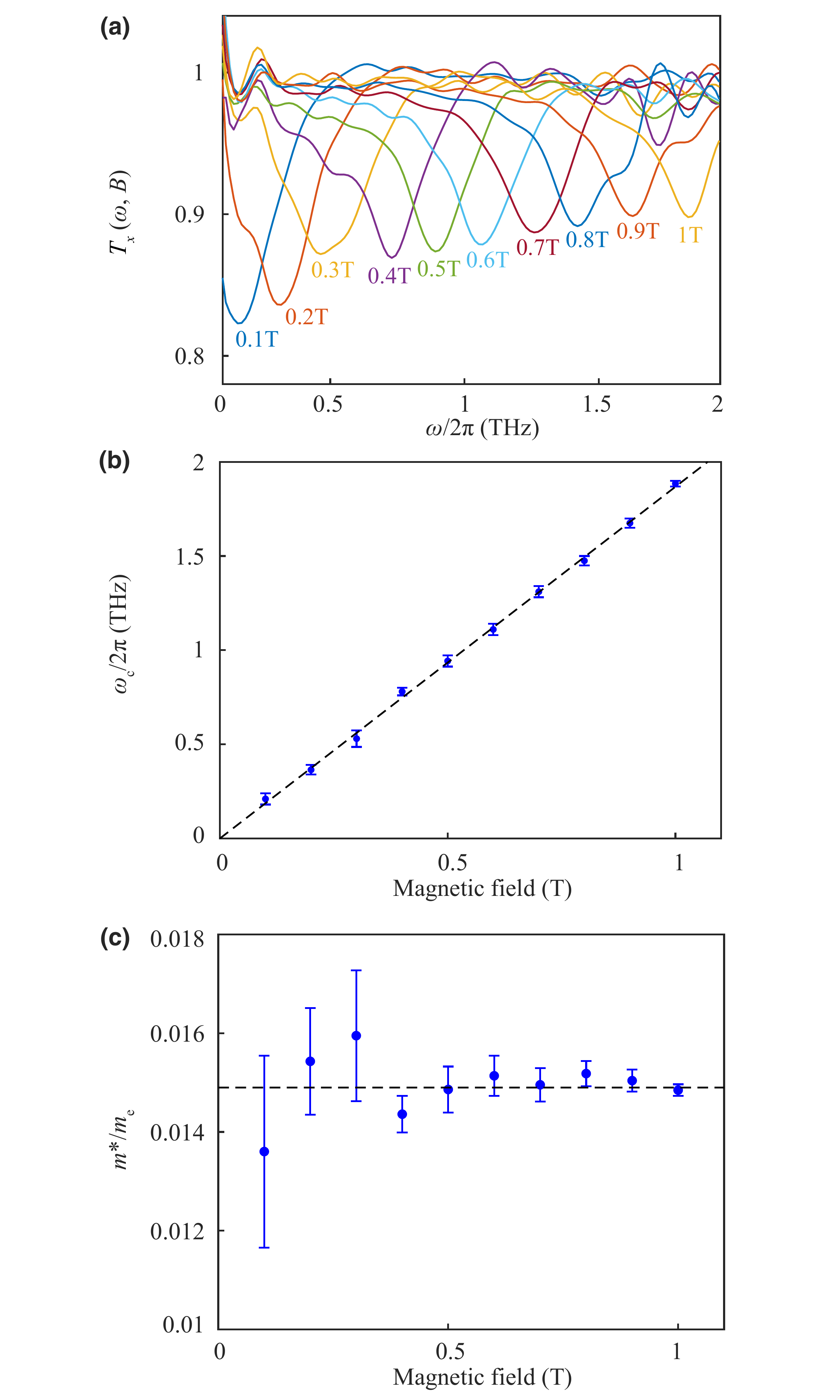}
	\caption{(a) THz transmission spectra of InSb measured at 200\,ps after photoexcitation at fluence 5.4\,nJcm$^{-2}$ with magnetic field ranging from 0.1 to 1\,T at a tempertaure of 4\,K. The small shoulders shown at high frequency ($1.5-2$\,THz) are experimental artefacts associated with the reduced spectral response of the measurement system over this range. (b) Cyclotron resonance frequency as a function of magnetic field, where the cyclotron resonance frequency corresponds to the minima of the transmission spectra shown in panel (a). The dotted line corresponds to a linear fit of the cyclotron frequency as a function of magnetic field. (c) InSb electron effective mass at 4\,K extracted from the cyclotron resonance frequency according to $\omega_{\rm c}=eB/m^*$. The dotted line corresponds to the effective mass obtained from the slope of the linear fit determined in panel (b).}
	\label{figure:cr_color}
\end{figure}

The THz transmission spectra of  InSb  photoexcited at a fluence of 5.4\fluence\ within a range of magnetic fields between 0.1 and 1\,T are displayed in Fig.~\ref{figure:cr_color}(a). The minimum of each THz transmission spectrum corresponds to the cyclotron resonance frequency \wc. As the magnetic field increases, \wc\ shifts to higher frequencies, which is expected from $\omega_{\rm c}=eB/m^*$. Owing to the extremely small bandgap of InSb, the Landau level energy is generally expected to increase sub-linearly with magnetic field~\cite{bowers1959magnetic}. However, since the applied magnetic field was only up to 1T, the relationship between the Landau level energy and the magnetic field is approximately linear as shown in Fig.~\ref{figure:cr_color}(b) where the dashed straight line represents a linear fit to the experimental data with $\omega_{\rm c}=0$ at $B=0$. The effective mass is then calculated according to $\omega_{\rm c}=eB/m^*$, which is constant at magnetic field below 1\,T as shown in Fig.~\ref{figure:cr_color}(c). The uncertainties of \wc\ shown in Fig.~\ref{figure:cr_color}(b) were determined by the frequency range within which the magnitude of cyclotron resonance in the transmission spectrum drops by 3\%. Correspondingly, the uncertainties of the effective mass shown in Fig.~\ref{figure:cr_color}(c) were calculated according to $\omega_{\rm c}=eB/m^*$. At small magnetic field ($B<0.3$\,T), since the cyclotron resonance frequency is close to the lower bound of the THz detection range ($f<0.5$\,THz), larger uncertainties in $\omega_{\rm c}$ are observed and therefore, the corresponding effective masses shown in Fig.~\ref{figure:cr_color}(c) have much larger error bars at $B=0.1$--0.3\,T. The horizontal dashed line in Fig.~\ref{figure:cr_color}(c) suggests that the electron effective mass of InSb is approximately 0.0149\me, which is consistent with previously reported value~\cite{dresselhaus1955cyclotron} within experimental error.

\section{Hot electron cooling in InSb} 
\begin{figure*}[ht!]
	\includegraphics[width=\textwidth]{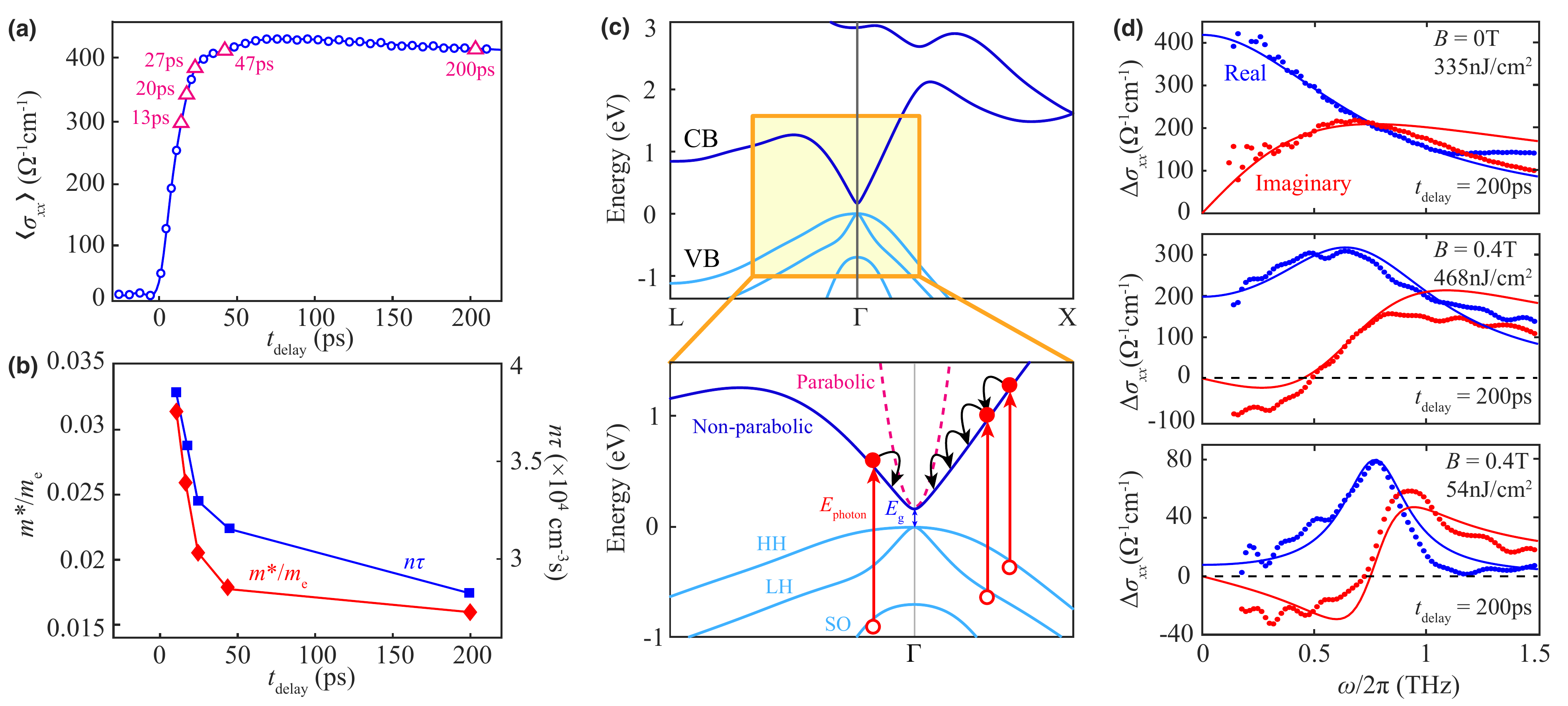}
	\caption{(a) Frequency-averaged photoconductivity  $\langle \sigma_{xx}\rangle$ of InSb measured at 4\,K as a function of time after photoexcitation by a 35\,fs,  335\fluence\ pulse of 1.55\,eV photons. The pink triangles represent the pump-probe delay times when the cyclotron resonance measurements were recorded. (b) Electron effective masses (red diamonds) calculated from the cyclotron resonance measurements which were performed at 13, 20, 27, 47 and 200\,ps after photoexcitation. The blue squares represent the product of charge-carrier density and relaxation time, $n\tau$. (c) Quasiparticle band structure of InSb, calculated within the $G_0W_0$ approximation (see Appendix E for details). The dark blue lines represent the conduction band (CB) and the light blue lines represent the valence band (VB), including the heavy hole (HH), light hole (LH) and split-off (SO) bands. The bottom panel is a zoom-in of the area near the $\Gamma$ valley. The pink dashed line represents a parabolic CB for comparison which overlaps with the non-parabolic CB near the $\Gamma$ minimum. The solid and open red circles represent an electron-hole pair generated by photoexcitation. The black arrows represent the hot electron cooling process. (d) Photoconductivity measured at 200\,ps after photoexcitation. The top panel shows $\Delta\sigma_{xx}$ measured without magnetic field, which is fitted with the Drude model given by Eq.~\eqref{eq:drude}. The middle and bottom panels show $\Delta\sigma_{xx}$ measured at magnetic field $B=0.4$\,T under fluence 468\fluence\ and 54\fluence\ respectively, fitted with the theoretical magneto-conductivity given by Eq.~\eqref{magneto_th}. }
	\label{figure:diff_delay}
\end{figure*}

We now examine the cooling of hot photoexcited electrons in InSb. Generally, the hot-carrier dynamics can be investigated via OPTPS by measuring the THz photoconductivity of the material at various pump-probe delay times, $t_\mathrm{delay}$, and by fitting a Drude-Lorentz model to each photoconductivity spectrum, the carrier momentum scattering time can be extracted as a function of $t_\mathrm{delay}$, which is used as a key parameter for analyzing hot-carrier cooling process in narrow-gap semiconductors such as GaAs and InAs~\cite{lloyd2012generalized}. This technique has also been applied to metal-halide perovskites whose dominant hot-carrier relaxation mechanism is found to be the electron-phonon interaction~\cite{monti2018efficient}. 
Here, time-resolved magneto-THz spectroscopy is utilized to probe the cyclotron effective mass of InSb at various delay times, which is another indicator of the hot-electron cooling process. Fig.~\ref{figure:diff_delay}(a) shows the frequency-averaged THz photoconductivity of InSb $\langle \Delta\sigma_{xx}\rangle$ as a function of time after photoexcitation with 1.55\,eV photons in the absence of magnetic field. While absorption occurs on the 35\,fs time scale, photoconductivity can be seen to rise on a timescale of tens of picoseconds. Time-resolved effective mass measurements conclusively show that this slow rise in $\langle \Delta\sigma_{xx}\rangle$ is a direct result of electron cooling (Fig.~\ref{figure:diff_delay}b); the effective mass reduces on the same timescale as the photoconductivity rises, in close agreement with Eq.~\eqref{eq:drude}. Although previous studies have reported THz pulse induced impact ionization and redistribution of electrons between the $\Gamma$ and L valleys~\cite{avsmontas2015monte,avsmontas2020impact}, the effective masses detected in our THz range suggest that our measurements are only sensitive to the electrons near the $\Gamma$ valley~\cite{kim2009accurate}. Therefore, our photoconductivity measurements are probing the $\Gamma$ valley non-parabolicity exclusively.

We sketch this hot-electron cooling process in a band structure picture in Fig.~\ref{figure:diff_delay}(c), the energy of the photo-injected electrons ($E_\mathrm{photon}$) is much larger than the bandgap ($E_\mathrm{g}$), which causes the hot electrons to reach the non-parabolic region of the conduction band. The change in effective mass is a result of hot electrons gradually cooling from this non-parabolic region towards the $\Gamma$ minimum where the conduction band is approximately parabolic and the electron effective mass is around 0.013\me~\cite{dresselhaus1955cyclotron}. Within the picosecond timescale of our experiment, we observed that at 200\,ps after photoexcitation, although the hot electrons cooled down to a much lower energy than the initially photoexcited energy state, they have an energy distribution that is still above the $\Gamma$ minimum, which leads to a larger value of effective mass (0.017\me).

The top panel in Fig.~\ref{figure:diff_delay}(d) shows the photoconductivity spectrum measured at 200\,ps after photoexcitation without magnetic field. The data agree fairly well with the Drude model, with the solid lines being a fit to Eq.~\eqref{eq:drude}. From the fit we extract momentum scattering rate $\gamma=4.7\times10^{12}$\,rad\,s$^{-1}$ (i.e. momentum scattering time $\tau=0.21$\,ps) and photo-injected electron density $n=7.1\times10^{15}$\den. This type of measurement is possible at any time after photoexcitation, and combined with the cyclotron data provides a complete picture of charge-carrier dynamics in a photoexcited semiconductor. 

On the other hand, the middle and bottom panels in Fig.~\ref{figure:diff_delay}(d) show $\Delta\sigma_{xx}$ measured under a magnetic field $B=0.4$\,T at fluence 468\fluence\ and 54\fluence\ respectively, both displaying a clear cyclotron resonance peak in the real part. In the Drude approximation, the photoconductivity spectra measured in magnetic field can be described by the magneto-conductivity tensor~\cite{lloyd2014terahertz},
{\small\begin{equation}\label{magneto_th} 
	\Delta\sigma_{ij}(\omega) = 
	\frac{\sigma_0}{(1-i\omega\tau)^2+(\omega_{\rm c}\tau)^2}
	\begin{bmatrix}
	1-i\omega\tau         & -\omega_\mathrm{c}\tau\\
	\omega_\mathrm{c}\tau & 1-i\omega\tau
	\end{bmatrix},
\end{equation}}where $i,j = x,y$, indexing the Cartesian directions in the plane perpendicular to the magnetic field. Here, $\Delta\sigma_{xx}$ represents the conductivity component measured in the direction parallel to the electric field of the incident THz pulse, which we label as the $x$-direction in Fig.~\ref{figure:cr_3d}(a). The off-diagonal components $\Delta\sigma_{xy}$ and $\Delta\sigma_{yx}$ represent the conductivity components which are orthogonal to the THz electric field. $\sigma_0=ne\mu$ represents the DC conductivity where $\mu$ is the electrical mobility of the charge carrier. By fitting $\Delta\sigma_{xx}$ given by Eq.~\eqref{magneto_th} to the experimental data in Fig.~\ref{figure:diff_delay}(d), the electron momentum scattering times at high fluence (468\fluence) and low fluence (54\fluence) are 0.3 and 0.9\,ps respectively. This is consistent with the fact that higher electron density results in more frequent electron-electron scattering and hence a shorter electron momentum scattering time.

The product of electron density and scattering time, $n\tau$, is shown in Fig.~\ref{figure:diff_delay}(b). It reduces rapidly over the initial 50\,ps, primarily as a result of $n$ decaying via bimolecular and Auger recombination. This should act to reduce photoconductivity, however the competing strong reduction in $m^*$ from 0.032 to 0.017\me\ has a much larger impact on $\Delta\sigma$ thus causing the observed photoconductivity to increase.

The photoconductivity spectrum of InSb was measured at 13, 20, 27, 47 and 200\,ps after photoexcitation and fitted with $\Delta\sigma_{xx}$ component of Eq.~\eqref{magneto_th} (see Fig.~\ref{delay} in Appendix B). Compared to the photoconductivity measurement shown in the top panel of Fig.~\ref{figure:diff_delay}(d), $\tau$ extracted from the cyclotron resonance spectra at 200\,ps increases by 1.8 times to 0.37\,ps, which is consistent with a restriction of phase space due to Landau quantization induced by magnetic field reducing the probability of electron scattering.

\section{Conduction band non-parabolicity probed via the quasi-Fermi level} 
\begin{figure}[ht!]
	\includegraphics[width=0.45\textwidth]{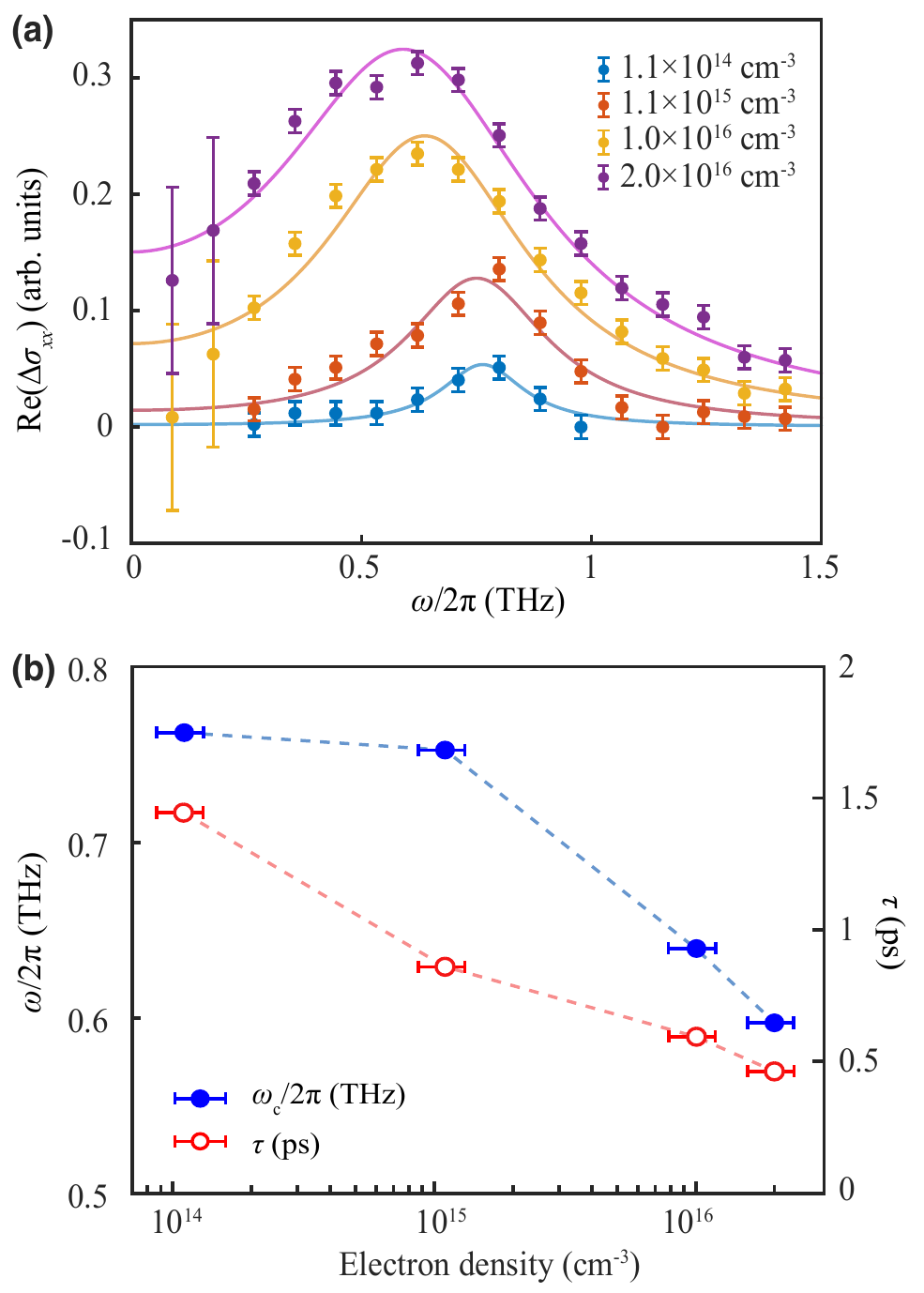}
	\caption{(a) Experimental cyclotron resonance spectra (solid circles) of InSb at a temperature of 4\,K measured 200\,ps after photoexcitation with  electron densities of $1.1\times10^{14}$, $1.1\times10^{15}$, $1.0\times10^{16}$ and $2.0\times10^{16}$\den\ under a magnetic field $B=0.4$\,T.   The uncertainties in Re$(\Delta\sigma_{xx})$ are obtained from the THz spectral response (see Appendix D). The solid lines represent the theoretical magneto-conductivity spectra which were obtained from Eq.~\eqref{eq:sigma0} and normalized to the experimental data. (b) Cyclotron resonance frequency, $f_{\rm c}=\omega_{\rm c}/2\pi$, (solid blue circles) and momentum scattering time, $\tau$, (open red circles) as a function of electron density.}
	\label{figure:cr_dist}
\end{figure}

Now we are in a position to investigate the non-parabolic feature of the conduction band of intrinsic InSb in more depth. Firstly, we measure the cyclotron resonance spectrum at regions of the conduction band above the $\Gamma$ minimum by changing the photoexcitation fluence. This is equivalent to the conventional method of tuning the Fermi energy level by implementing different doping density in n-type InSb but with better flexibility and consistency since only one intrinsic InSb sample was used. We measured the cyclotron resonance of InSb at 200\,ps after photoexcitation when the electrons have thermalized and sufficiently cooled. By considering the effect of charge-carrier diffusion, the photoexcitation fluence spanning over two orders of magnitude is converted to electron density (details of electron density calculation are provided in Appendix C). The resultant electron density ranges from $1.1\times10^{14}$ to $2.0\times10^{16}$\den, which covers most of the electron density range of previously studied n-type InSb~\cite{infra_burstein1956cyclotron,infra_keyes1956infrared,infra_palik1961infrared,combine_mccombe1969infrared,smith1959energy,johnson1970infrared}. 
To correlate the experimental results with the electronic band non-parabolicity, we calculate the magneto-conductivity component, $\Delta\sigma_{xx}$, by solving the Boltzmann transport equation in the constant relaxation time approximation~\cite{zawadzki1974electron,ziman2001electrons}. Using the analytical Kane model expression for the conduction band dispersion, the DC conductivity in Eq.~\eqref{magneto_th} is expressed as, 
\begin{equation}
\small{
\displaystyle{\sigma_0 = -\frac{4e^2\tau}{3\pi^2\hbar^2}\sqrt{\frac{2m_0^*}{\hbar^2}}\int_0^\infty \frac{[E(1+\frac{E}{E_g})]^{3/2}}{1+2\frac{E}{E_g}}\frac{\partial f(E,\mu_n,T)}{\partial E}{\rm d}E}}, 
\label{eq:sigma0}
\end{equation}
where $f(E,\mu_n,T)$ is the Fermi-Dirac distribution calculated at a given electronic temperature. For the effective mass at the conduction band minimum, $m_0^*$, we use the calculated effective mass obtained with two methods: from the numerical second derivative of the \emph{ab initio} quasiparticle energies and from fitting the conduction band using the Kane model. Both methods yield the same value of 0.013\me, in close agreement with experiment, further validating our computational approach. Furthermore, we explicitly verify the accuracy of the Kane model for InSb through the direct comparison between the conduction band dispersion calculated using the Kane model and the {\it ab initio} $G_0W_0$ approximation (see Appendix E). In all our calculations we determine the chemical potential $\mu_{n}$ by equating the calculated charge carrier concentration, $n = \int_0^\infty g(E) f(E,\mu_n,T) dE$, with the value measured experimentally where $g(E)$ is the conduction band density of states calculated using the Kane model as $g(E)=\frac{1}{2\pi^2}\left(\frac{2m_0^*}{\hbar^2}\right)^{3/2}(1+2E/E_g)\sqrt{E(1+E/E_g)}$. The temperature $T$ is chosen so as to obtain the best alignment of the calculated resonance peak to the experimental measurements, and the experimentally measured relaxation time $\tau$ is used in all calculations. Fig.~\ref{figure:cr_dist}(a) shows the experimentally measured cyclotron resonance spectra (solid circles) in comparison with the theoretical results (solid lines). The magnitudes of the calculated cyclotron resonance spectra have been normalized to those of the experimental data.
The uncertainty increases at low frequencies ($f<0.3$\,THz) owing to reduced sensitivity of the THz spectrometer there and systematic errors (see Appendix D). Therefore, the calculated results agree with the measurements within experimental uncertainties. A summary of the measured cyclotron frequency and momentum scattering time of the resonances is provided in Fig.~\ref{figure:cr_dist}(b) as a function of electron density. It can be seen that $\tau$ decreases significantly with increasing electron density. 
This trend is attributed to the increasing electron-electron scattering and increasing range of effective mass values spanned by the electron population with increasing electron density.

\begin{figure}[ht!]
	\includegraphics[width=0.48\textwidth]{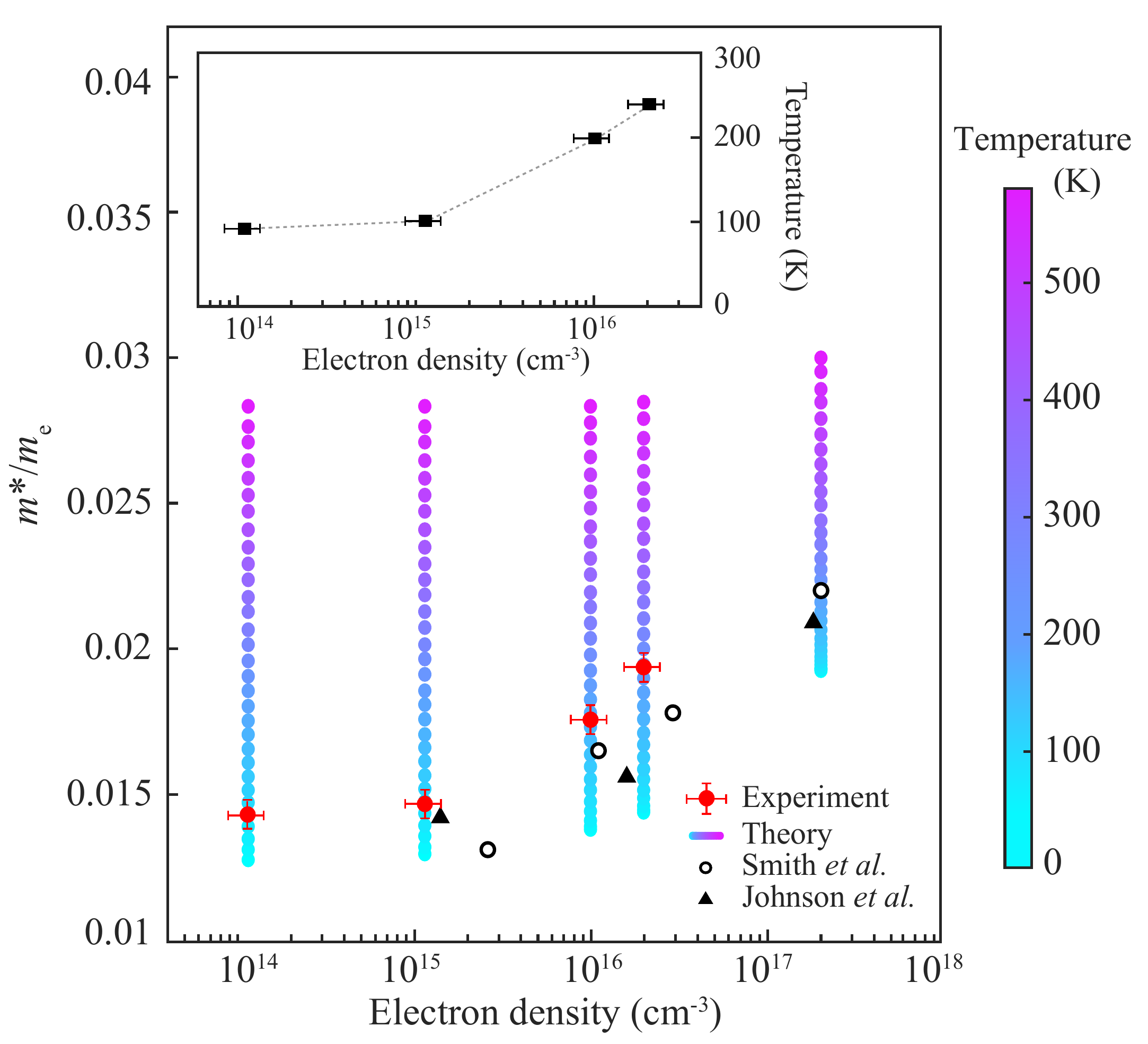}
	\caption{InSb electron effective mass as a function of electron density. The red solid circles with error bars represent the experimental effective mass measured in a cryostat held at 4\,K for photoinjected electron densities $1.1\times10^{14}$, $1.1\times10^{15}$, $1.0\times10^{16}$ and $2.0\times10^{16}$\den. The blue-purple colored solid circles represent the theoretically calculated effective masses, with the electron temperature indicated by the color scale on the right. The open circles and solid triangles represent previously reported effective masses measured in n-type InSb with various doping densities~\cite{smith1959energy,johnson1970infrared}. The inset shows the electronic temperature as a function of electron density. The error bars are defined in Appendix D.}
	\label{figure:meff_temp}
      \end{figure}

In Fig.~\ref{figure:meff_temp} we compare theoretical effective masses calculated for a range of electron temperatures between 1 and 580 K, and various charge carrier concentrations, with corresponding experimental data. We observe very similar trends from both theory and experiment, whereby the cyclotron effective masses increases with increasing charge carrier concentration, as a consequence of the non-parabolic conduction band. 
Furthermore, the electronic temperature for which theoretical and experimental effective masses coincide increasing with carrier concentration (see the inset), follows the intuition expected from Fermi-Dirac statistics. It should be noted that this is a rough approximation of the electronic temperature. In our calculations, the electronic temperature and the chemical potential were used interactively via Fermi-Dirac distribution to determine the electron effective mass. As the electron density increases, the quasi-Fermi level (i.e. the chemical potential) shifts to higher energy level in the conduction band due to the effect of Pauli blocking of the conduction band states. By taking this into account, our calculations show that the electron effective mass increases monotonically with the electronic temperature (see Fig.~\ref{mass_temp} in Appendix E). Compared to a previous study where the transient absorption and a non-linear cross-correlation technique were used for inferring the charge-carrier cooling rate~\cite{lobad2004carrier}, the electronic temperature extracted from our study is surprisingly high. We attribute this difference to a few reasons. Firstly, in our experiment, the electrons were photoexcited with 1.55\,eV photons, which gave a much higher initial electron temperature ($\sim10^4$\,K), whereas in the other study, charge carriers were injected with only 650\,K excess energy. Therefore, the charge-carrier cooling process is likely to be different and not necessarily comparable between those two studies. Secondly, as shown in Fig.~\ref{figure:diff_delay}(a), the photoconductivity rises slowly after the initial photo-injection, which reaches its peak only after about 50\,ps. This indicates a rather slow injections of hot electrons and may prolong the electron thermalization process. Apart from that, non-equilibrium phonon reabsorption may have also contributed to this elevated electronic temperature, which however, will require another study to investigate. 

Overall, we note that the calculated trend shown in Fig.~\ref{figure:meff_temp} agrees with both our experimental data and other prior measurements (under different experimental conditions), thereby unifying the general trend for the increasing cyclotron effective mass with increasing number of electrons that populate the non-parabolic conduction band.
 

\section{Conclusion}
In summary, we have shown that time-resolved magneto-THz spectroscopy is  a versatile technique for probing the dynamics of hot charge carriers in semiconductors. It allows electronic properties including charge density, scattering rate, mobility and cyclotron effective mass to be measured on a picosecond time scale. We calculated the quasiparticle band structure of InSb within the $G_0W_0$ approximation and found excellent agreement with THz experiments. Conduction band non-parabolicity was experimentally observed by photo-doping InSb in close agreement with theoretical predictions. Hot electrons were injected into InSb by photoexcitation, and hot carrier cooling was observed over a period of 200\,ps. The ability to measure simultaneously effective mass and conductivity on this timescale is particularly significant. Furthermore the technique can be extended to observe sub-picosecond carrier dynamics and is applicable to a wide range of semiconductors and semiconductor heterostructures. Recent achievements in polarization-sensitive THz detectors~\cite{peng2020three} will make this technique more accessible.

\begin{acknowledgements}
	We acknowledge funding provided by the Engineering and Physical Sciences Research Council (EPSRC). MBJ thanks the Alexander von Humboldt Foundation for support. We thank Dr.~Mark Ashwin (University of Warwick) for growing the InSb sample. MRF would like to acknowledge the support of the John Fell Oxford University Press (OUP) Research Fund, and the use of the University of Oxford Advanced Research Computing (ARC) facility in carrying out this work~\cite{Richards2015fund}.
\end{acknowledgements}

\section*{Appendix A: Polarization-resolved magneto-THz spectroscopy}
\begin{figure*}[ht!]
	\includegraphics[width=\textwidth]{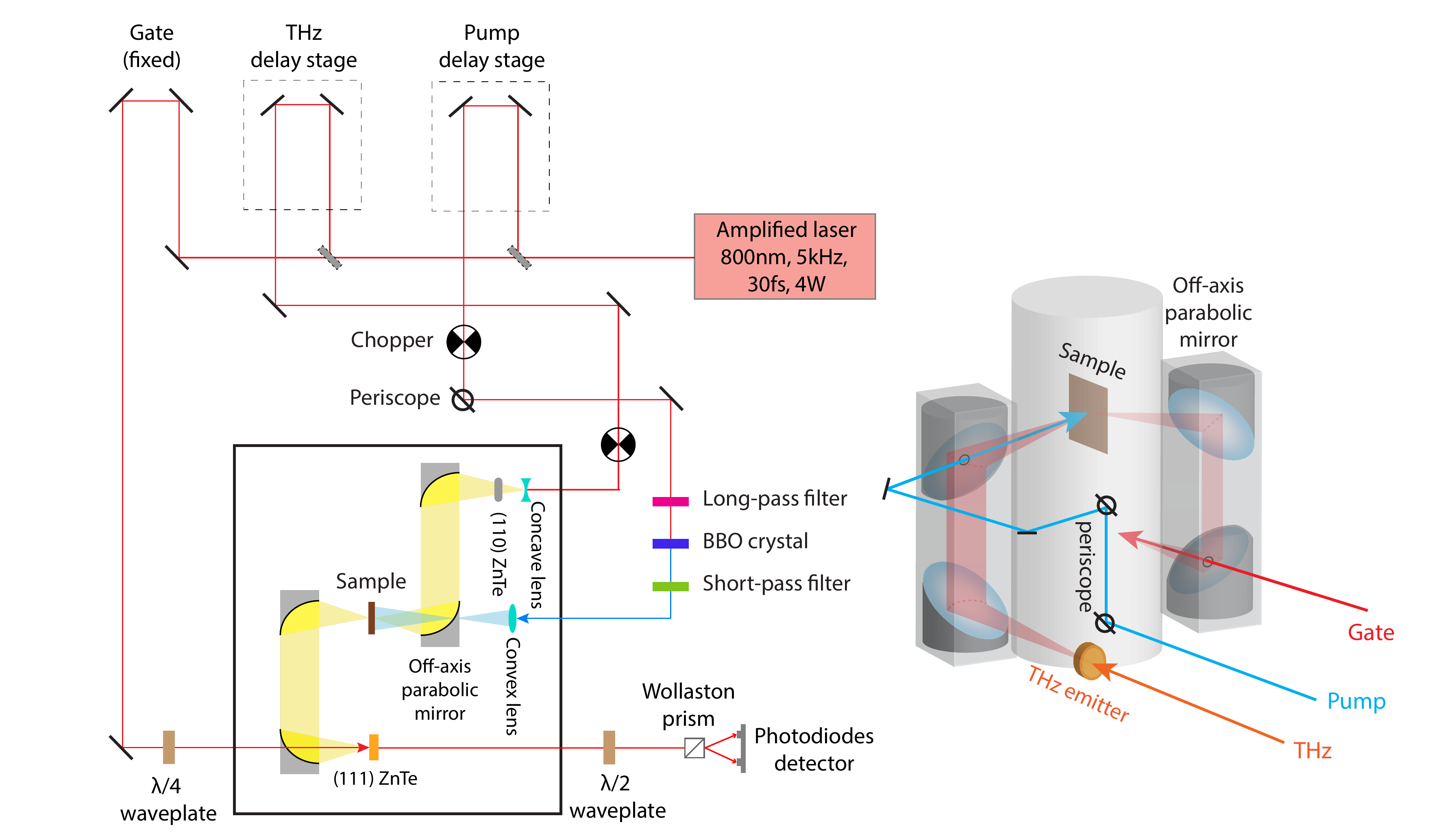}
	\caption{Schematic for magneto-THz spectroscopy setup. The cartoon on the righthand side shows the configuration of the superconducting magnet along with two pairs of off-axis parabolic mirrors which are used for focusing the THz pulse onto the sample and the (111) ZnTe detection crystal respectively.}
	\label{figure:shcematic}
\end{figure*}
We used time-resolved magneto-THz spectroscopy to study the cyclotron resonance phenomenon in InSb. Fig.~\ref{figure:shcematic} shows a schematic diagram of the experimental setup. An amplified ultrafast (35\,fs) laser pulse with central wavelength at 800\,nm was split into three beams: a ``gate beam'', a ``THz beam'' and a ``pump beam''. The ``THz beam'' was used to generate a single-cycle THz pulse via optical rectification in a 1\,mm-thick (110) ZnTe crystal.  The ``gate beam'' was used to detect the THz signal via electro-optic sampling, which used a 1\,mm-thick (111) ZnTe crystal, a quarter-wave plate, a half-wave plate, a Wollaston prism and a pair of balanced photodiodes. The ``pump beam'' with central wavelength at 800\,nm was used for photoexciting the intrinsic InSb sample. Photoecitation of the InSb generated free electrons its conduction band and thus reduced the transmission of the THz pulse through it. This is the so-called optical-pump-THz-probe spectroscopy (OPTPS), which has been widely used for investigating semiconductor materials' optoelectronic properties, such as charge-carrier mobility and relaxation dynamics~\cite{mobility_THz_christian2,johnston2016hybrid,lui2001ultrafast,wang2004electronic}. Our magneto-THz spectroscopy is an extension of this OPTPS where an external magnetic field generated by a superconducting magnet (Oxford Instruments) is applied perpendicularly to the sample surface, which induces cyclotron motion of the photoinduced charge carriers and leads to cyclotron resonance absorption of the THz pulse at certain frequency.

In order to measure the orthogonal polarization components of the transmitted THz pulse separately, we used a single (111) ZnTe crystal as the detection crystal whose detection efficiencies in orthogonal orientations are similar, as opposed to the conventionally used (110) ZnTe which has a preferred detection orientation depending on the polarization of the incident light. The voltage difference, $\Delta$PD, measured by the pair of photodiodes was proportional to the electric field strength of the THz pulse, which is expressed as,
\begin{equation}
	\Delta\mathrm{PD}\propto{E}_{\bar{2}11}\sin(2\psi-4\delta)+E_{0\bar{1}1}\cos(2\psi-4\delta),
\end{equation}
where ${E}_{\bar{2}11}$ and $E_{0\bar{1}1}$ represent the THz electric field along the $\langle\bar{2}11\rangle$ and $\langle0\bar{1}1\rangle$ axes of the (111) ZnTe crystal, respectively. “$\psi$ is the angle between the Wollaston prism and the $\langle0\bar{1}1\rangle$ axis of the (111) ZnTe crystal. $\delta$ is the angle between the half-wave plate and the Wollaston prism. A certain linear polarization component of the THz pulse was selected by rotating the half-wave plate, with the quarter-wave plate used for balancing the photodiode at that specific polarization state.  This combinational use of a quarter-wave plate and a half-wave plate enabled the  measurement of orthogonal polarizations without rotating the ZnTe crystal or the Wollaston prism, while maintaining similar detection sensitivity in both polarizations~\cite{van2005EOSpol}.

\section*{Appendix B: Cyclotron resonance measured at various delay times}
\begin{figure}[ht!]
\includegraphics[width=0.5\textwidth]{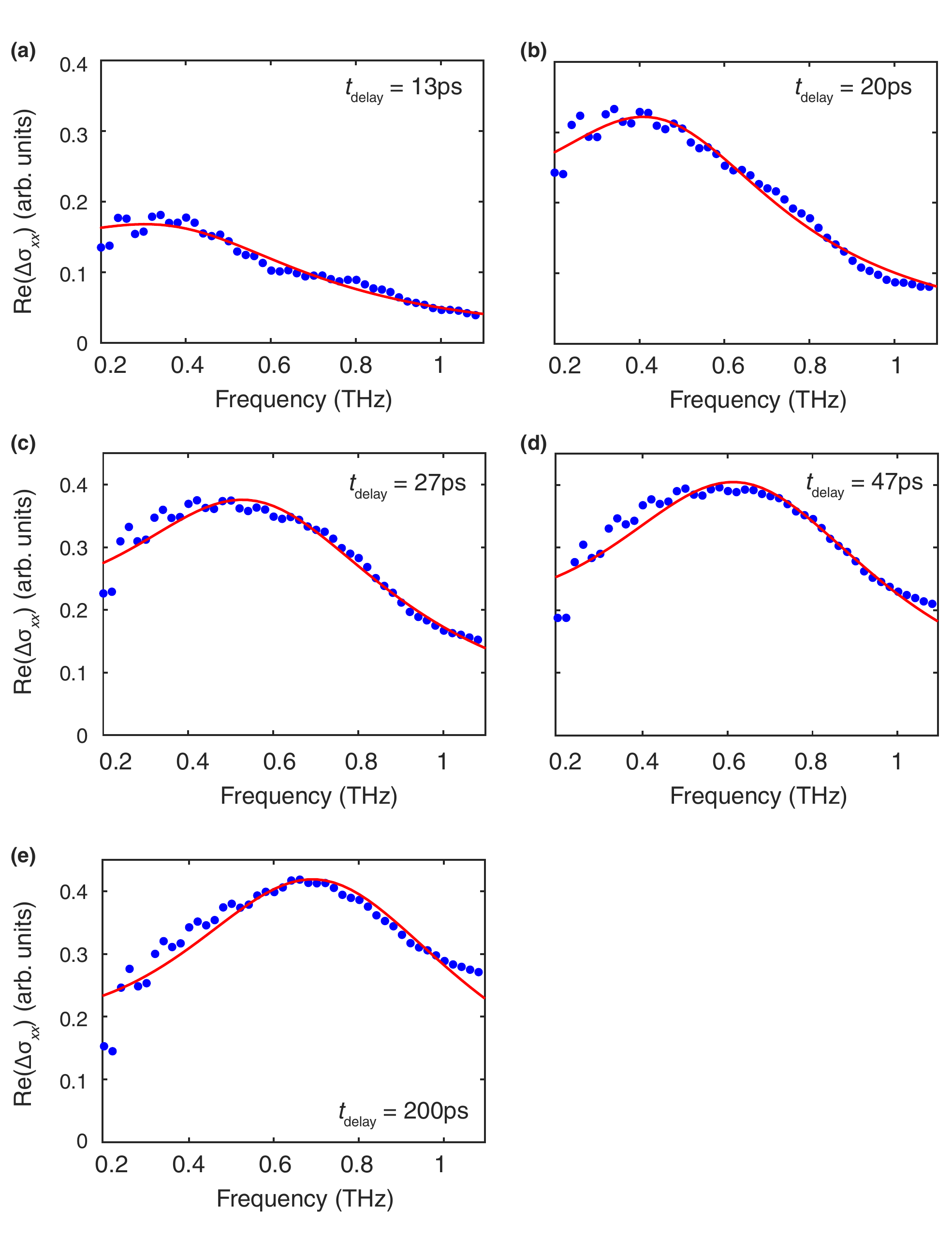}
\centering
\caption{Cyclotron resonance spectra of InSb measured at $B=0.4$\,T at (a) 13\,ps, (b) 20\,ps, (c) 27\,ps, (d) 47\,ps and (e) 200\,ps after photoexcitation under fluence 335\fluence. The blue solid circles represent the experimental data, whereas the red curves are magneto-conductivity fits given by the real part of $\Delta\sigma_{xx}$ component of Eq.~\eqref{magneto_th}.} 
\label{delay}
\end{figure}

Fig.~\ref{delay} shows the cyclotron resonance spectra of InSb measured at different delay times after photoexcitation, which were fitted by the $\Delta\sigma_{xx}$ component of the Drude magnetoconductivity tensor (Eq.~\ref{magneto_th}). The extracted cyclotron resonance frequencies are converted to electron effective masses and plotted in Fig.~\ref{figure:diff_delay}(b). Table~\ref{table1} lists the fitting parameters for the cyclotron resonance spectra shown Fig.~\ref{delay}. Similarly, we fitted the experimental data of Fig.~\ref{figure:cr_dist}(a) with the magneto-conductivity tensor ($\Delta\sigma_{xx}$) to extract the cyclotron resonance frequency and electron momentum scattering time at different electron densities. Table~\ref{table2} lists the corresponding fitting parameters for Fig.~\ref{figure:cr_dist}(a), which were then used for calculating the theoretical curves obtained from the Kane model (Eq.~\ref{eq:sigma0}).

\begin{table}[ht!]
\begin{tabular}{cccccc}
\hline
$t_{\rm delay}$ (ps) & 13 & 20 & 27 & 47 & 200 \\
\hline
$f_{\rm c}(=\omega_{\rm c}/2\pi)$ (THz) & 0.356 & 0.432 & 0.545 & 0.626 & 0.701\\
$\tau$ (ps) & 0.389 & 0.427 & 0.385 & 0.382 & 0.372\\
\hline
\end{tabular}
\caption{Fitting parameters for the cyclotron resonance spectra measured at different delay times, $t_{\rm delay}$, shown in Fig.~\ref{delay}.}
\label{table1}
\end{table}

\begin{table}[ht!]
\begin{tabular}{ccccc}
\hline
$n$ (cm$^{-3}$) & $1.1\times10^{14}$ & $1.1\times10^{15}$ & $1.0\times10^{16}$ & $2.0\times10^{16}$\\
\hline
$f_{\rm c}$ (THz) & 0.763 & 0.753 & 0.640 & 0.597\\
$\tau$ (ps) & 1.445 & 0.859 & 0.594 & 0.462\\
\hline
\end{tabular}
\caption{Fitting parameters for the cyclotron resonance spectra measured at different electron densities, $n$, shown in Fig.~\ref{figure:cr_dist}(a).}.
\label{table2}
\end{table}

\section*{Appendix C: Estimation of electron density under photoexcitation}
In order to investigate the non-parabolic feature of InSb, the conduction band was probed by increasing photoexcitation fluence, which enables the electrons with higher density to reach higher energy level in the conduction band and hence exhibit larger effective mass. Knowing the photoexcitation fluence, the initial electron density, $n_{\rm I}$, was estimated according to the following equation,
\begin{equation}
	n_{\rm I}=\frac{\mathrm{Fluence\cdot\alpha}}{{E}_\mathrm{photon}},
	\label{eq:n}
\end{equation}
where $\alpha\approx10^5$\,cm$^{-1}$ is the absorption coefficient of InSb at 800\,nm~\cite{aspnes1983dielectric} and $E_\mathrm{photon}$ is the energy of a single photon at 800\,nm. The fluence was determined by the photoexcitation power ($P$) and the effective area ($A_{\rm eff}$) between the pump beam and the THz beam~\cite{mobility_THz_christian2}, which is expressed as ${\rm Fluence}=P/(5{\rm kHz}\cdot{A}_{\rm eff})$.

\begin{figure}[ht!]
\includegraphics[width=0.35\textwidth]{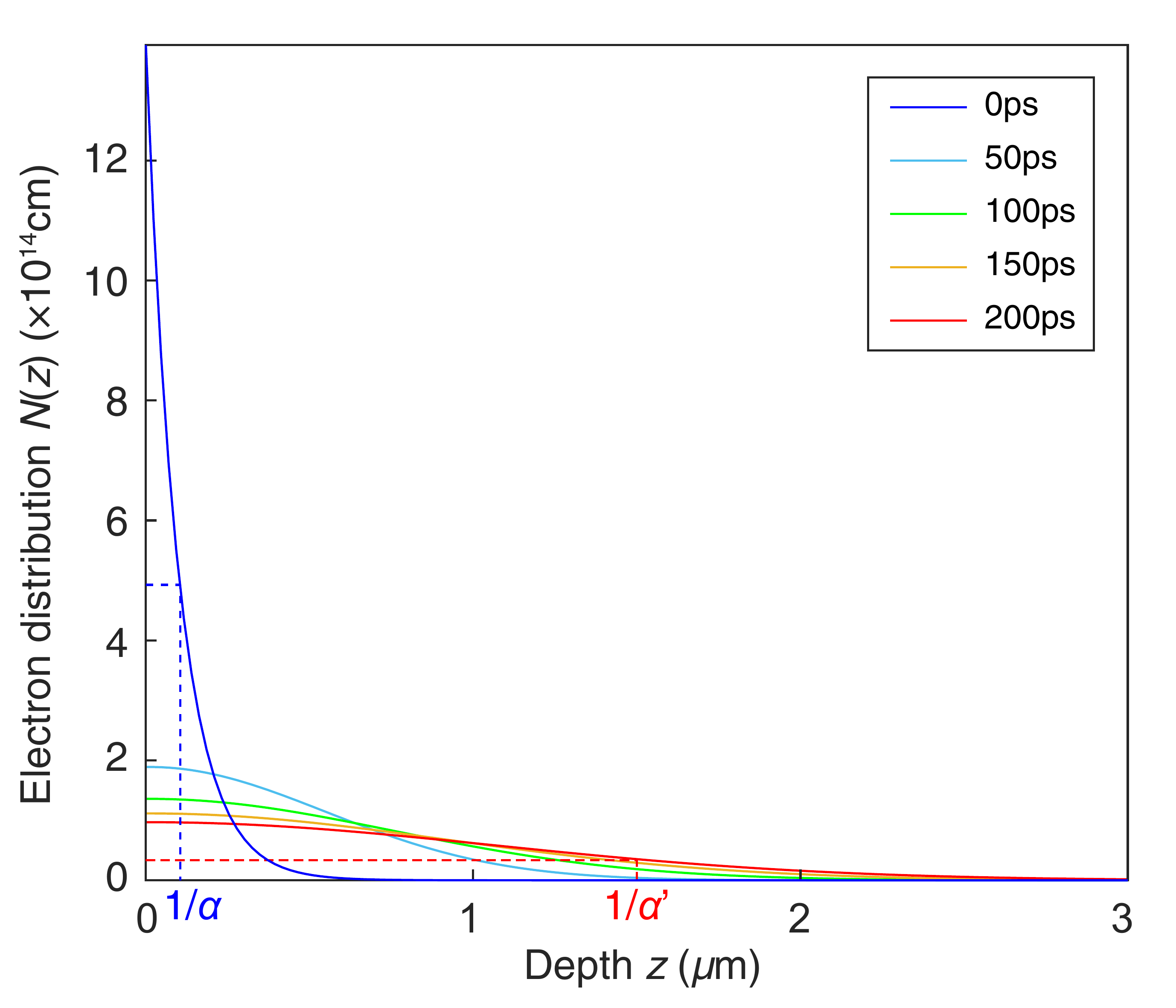}
\centering
\caption{Electron distribution profiles as a function of depth in InSb calculated at 0, 50, 100, 150 and 200\,ps after photoexcitation. The initial photoexcitation fluence is 5.4\fluence. The dashed lines represent the absorption depth $1/\alpha$ and effective absorption depth $1/\alpha'$ at 200\,ps.}
\label{fig:diff}
\end{figure}

Eq.~\eqref{eq:n} gives an estimation of the electron density at $t=0$\,ps assuming a uniform charge-carrier distribution. However, since the effective mass was measured at 200\,ps after photoexcitation, the electron density decreased significantly owing to charge-carrier diffusion. Therefore, to account for the effect of diffusion, we model the electron distribution $N(t,z)$ as a function of time $t$ and depth $z$ according to the one-dimensional diffusion equation,
\begin{equation}
\frac{\partial{N}}{\partial{t}}=D\frac{\partial^2{N}}{\partial{z^2}},
\label{eq:diff}
\end{equation}
where $D=\mu{k_{\rm B}}T/e$ is the diffusion coefficient determined by the electron mobility $\mu$ and temperature $T$. $t$ is the time period during which the electrons have diffused through the sample and $z$ is the distance travelled by the electrons in a direction perpendicular to the sample surface. The reduction of electron density owing to electron-hole recombination was neglected. Eq.~\eqref{eq:diff} was solved using finite-difference method. In Fig.~\ref{fig:diff} the electron distribution profile is plotted at different photoexcitation delay times, and can be seen to gradually flatten out at later times as a result of diffusion. 

The initial electron distribution profile at $t=0$\,ps is assumed as an exponential decay curve based on the Beer-Lambert law,
\begin{equation}
N(0,z)=N_0\exp(-z\cdot\alpha),
\end{equation}
where $N_0$ is the number of electrons at the sample surface which can be determined by the fluence, given the fact the total number of electrons $N(0,z)$ is equal to the number of photons $N_{\rm photon}$, i.e. $\int_0^\infty[{N_0}\exp(-z\cdot\alpha)]dz=N_{\rm photon}$.

To estimate the electron density at 200\,ps after photoexcitation, the effective absorption depth was taken as the depth at which the diffused electron population was $\exp(-1)$ of the surface electron density. As demonstrated in Fig.~\ref{fig:diff}, the effective absorption depth, $1/\alpha'$, at 200\,ps is much longer than that at 0\,ps, which is expected since the electrons diffuse further into the sample at later times. As a result, the electron density at 200\,ps was significantly reduced and was calculated as
\begin{equation}
n(t)=\frac{\int_0^{1/\alpha'}N(t,z)dz}{A_{\rm eff}(1/\alpha')}.
\label{n_diff}
\end{equation}

\section*{Appendix D: Experimental uncertainties associated with the cyclotron resonance measurements}
\subsection*{Error bars for electron densities and effective masses}
It should be noted that due to the experimental uncertainties of the measured spot sizes of the THz and optical pump beams (which are approximately $\pm0.2$\,mm), the effective area, $A_{\rm eff}$, used in Eq.~\eqref{n_diff} has an uncertainty of $\sim$13\%, which gives rise to the horizontal error bars of the electron densities shown in Fig.~\ref{figure:cr_dist}(b) and Fig.~\ref{figure:meff_temp}. As for the vertical error bars shown in Fig.~\ref{figure:meff_temp} for the effective masses, they are determined from the width of each cyclotron resonance spectrum measured at different electron densities. Specifically, we calculated the experimental uncertainties of \wc\ (i.e. $\Delta\omega_{\rm c}$) based on the width of the cyclotron resonance spectrum. $\Delta\omega_{\rm c}$ was determined such that within the range $(\omega_{\rm c}\pm\Delta\omega_{\rm c})$, the real part of the cyclotron resonance spectrum, Re$(\Delta\sigma_{xx})$ shown in Fig.~\ref{figure:cr_dist}(a), is greater than 97\% of its maximum. In other words, at $\omega=(\omega_{\rm c}+\Delta\omega_{\rm c})$ and $(\omega_{\rm c}-\Delta\omega_{\rm c})$, Re$(\Delta\sigma_{xx})$ drops by 3\%. According to $\omega_{\rm c}=eB/m^*$, the uncertainties for the effective masses are then determined by $\Delta\omega_{\rm c}$.

\subsection*{Spectral response of the THz spectrometer}
\begin{figure}[ht!]
\includegraphics[width=0.35\textwidth]{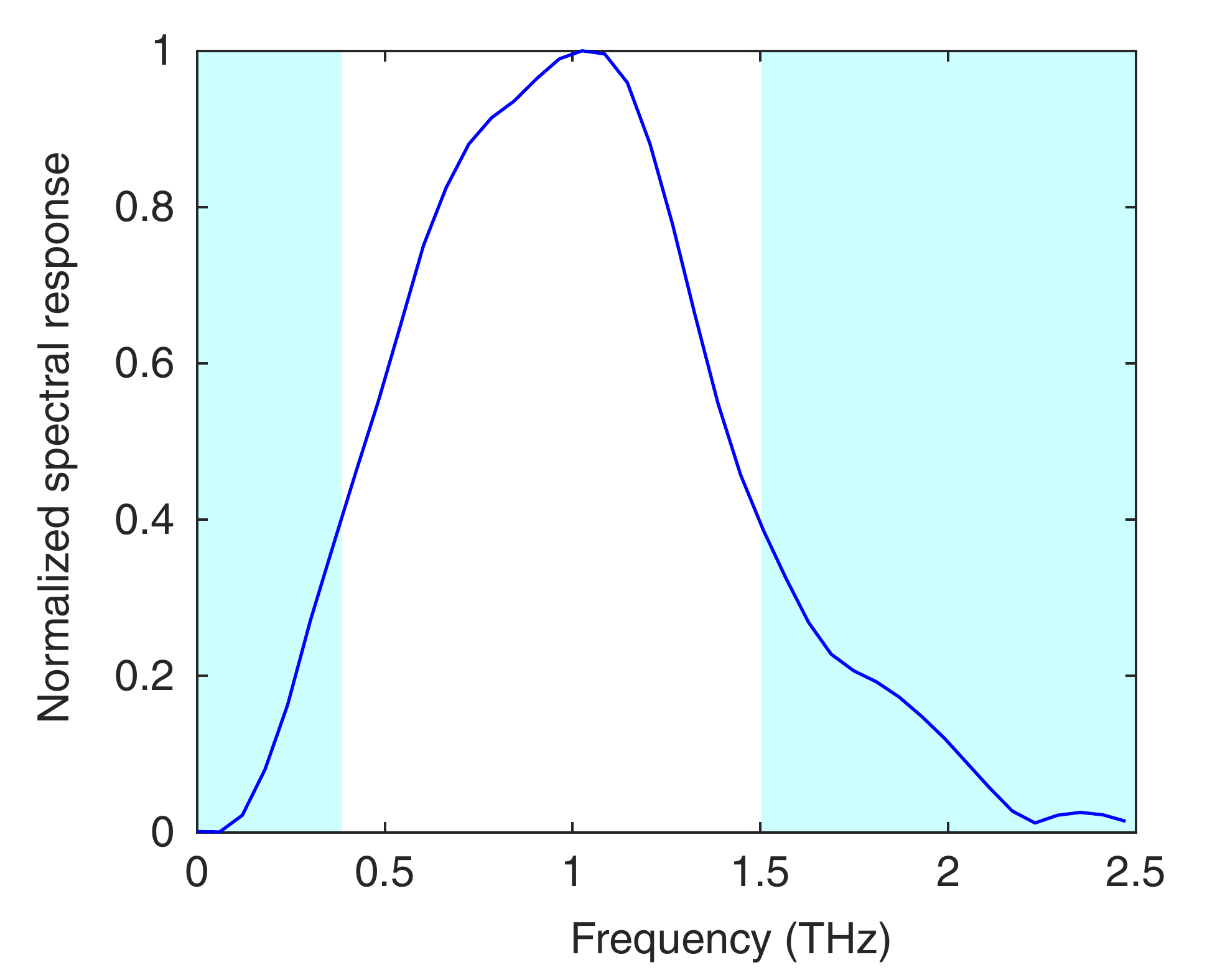}
\centering
\caption{Normalized raw spectrum of the THz electric field for the THz spectroscopy system used in this study without the application of a spectral response correction. The shaded regions indicate the frequencies at which the spectral amplitude falls to less than 40\% of the spectrometer’s peak spectral response. The signal-to-noise ratio and random errors are expected to be more significant in these regions.}
\label{response}
\end{figure}

The signal-to-noise ratio of a THz measurement generally varies through the spectrum as it is affected by the spectral response of the THz spectrometer. The spectrometer's spectral response takes into account the THz-frequency response of the THz emitter, THz detector and optical elements. Generally, the spectral response drops off at low and high frequencies, which leads to higher uncertainties in these spectra regions. Fig.~\ref{response} shows the raw (uncorrected) spectrum of the THz system used in this study without a sample. The frequency dependence of the error bars shown in Fig.~\ref{figure:cr_dist}(a) were calculated from the inverse of the spectral intensity and scaled by the standard deviation of three repeat measurements such as in Figure 1 of Ref.~\cite{davies2018impact}. Thus, the error bars presented in this study represent random errors, and it is expected that these errors will increase at low and high frequency owing to the lower signal-to-noise ratio in these regions. 

Notably, the experimental photoconductivity spectra shown in Fig.~\ref{figure:cr_dist}(a) exhibit some deviations from the theoretical curves at $f<0.3$\,THz.  At these frequencies random errors, as indicated by the error bars, are particularly large owing to the poor spectral response of the experiment at these frequencies.  In addition to random errors, three of the four data sets presented in Fig.~\ref{figure:cr_dist}(a) appear to underestimate the theoretical conductivity at $f<0.3$\,THz.  This may indicate the presence of small systematic errors or an underestimate of the momentum scattering rate in the theoretical model.  A potential source of systematic error in the experiment is the imperfect alignment of the spectrometer, and beam aperturing.  In particular, a slight ellipticity of the reference THz pulse is apparent in Fig.~\ref{figure:cr_3d}(b).  Our conversion of the raw THz data to conductivity assumed this pulse was linearly polarised, as for that scan (InSb at 4K in the dark) free charges should be absent, thus a small systematic error in the displayed $\sigma_{xx}$ values is possible.  Alternatively, it is possible that in our model, we are underestimating the momentum scattering time $\tau$ in the low frequency region. An increased value of $\tau$ would lower the value of $\sigma_{xx}$ at low frequency, however, it would also narrow the cyclotron resonance peaks in the region of high spectral response (0.3--1.5\,THz), leading to a larger deviation between experiment and theory in that spectral region. This effect can be seen clearly by the fitted magnetoconductivity in Fig~\ref{figure:diff_delay}(d).

\section*{Appendix E: Quasiparticle band structure calculation}
We calculated the quasiparticle band structure, $E_{n\mathbf{k}}$ from first principles, within the $G_0W_0$ approximation, as~\cite{hedin1965new, hybertsen1986electron}
\begin{equation}
E_{n\mathbf{k}} = \epsilon_{n\mathbf{k}} + Z_{n\mathbf{k}} \langle n\mathbf{k} | \Sigma - V_{\rm xc} | n\mathbf{k} \rangle,
\end{equation}
$\epsilon_{n\mathbf{k}}$ and $|n\mathbf{k}\rangle$ are mean-field energies and wave functions corresponding to the electronic band $n$ and wave vector $\mathbf{k}$, calculated within density functional theory (DFT)~\cite{hohenberg1964inhomogeneous}, $V_{\rm xc}$ is the exchange-correlation potential, $Z({\omega}) =\Big[1-{\rm Re}(\partial \Sigma / \partial \omega) \Big]^{-1}$ is the quasiparticle renormalization factor, and $\Sigma$ is the electron self energy, calculated in the $G_0W_0$ approximation as the convolution of the single-particle Green's function, $G_0$, and the screened Coulomb interaction, $W_0$.  

\begin{figure}[ht!]
\includegraphics[width=0.5\textwidth]{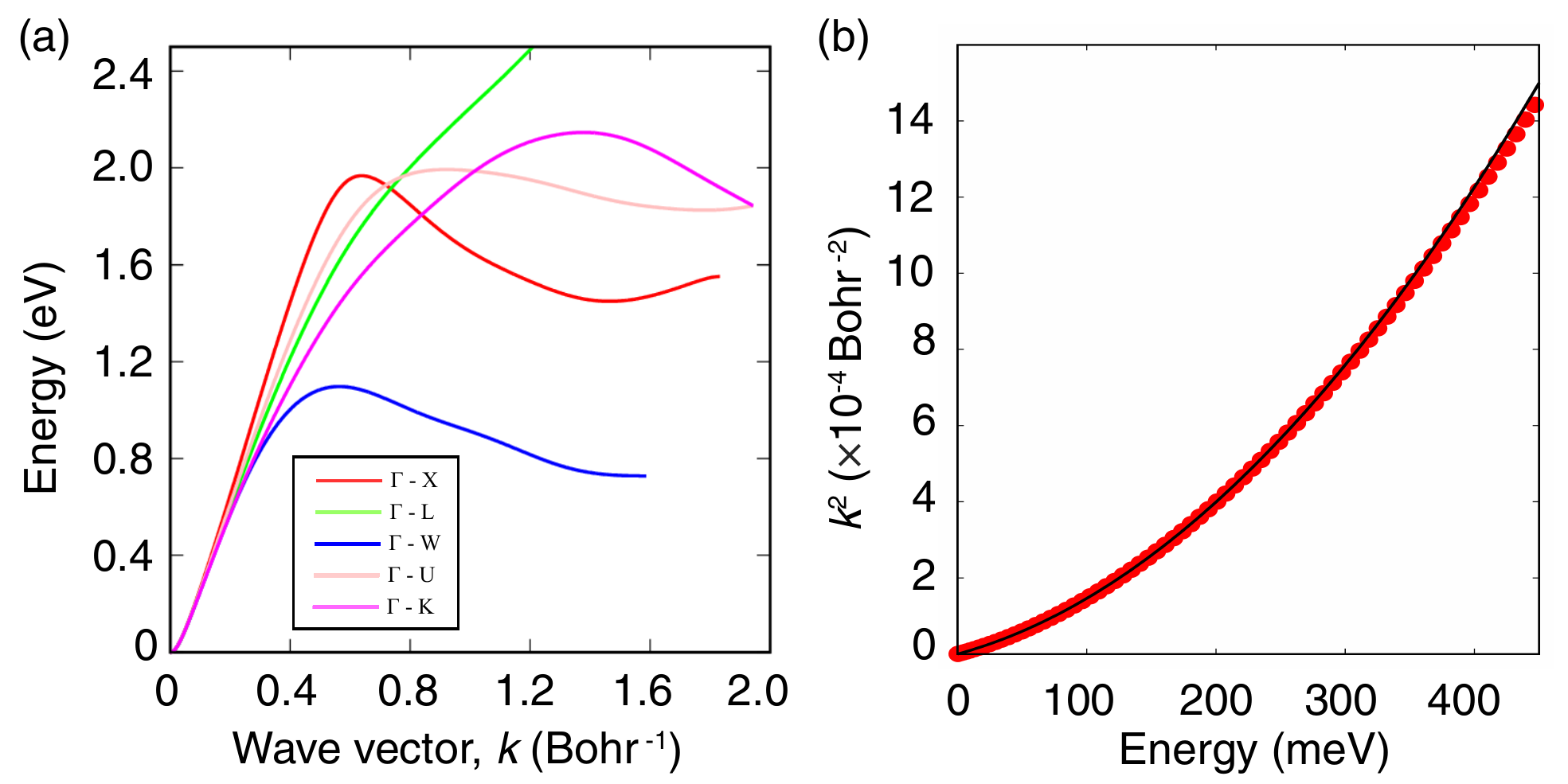}
\centering
\caption{(a) $G_0W_0$ conduction band energy of InSb as a function of wave vector along several high symmetry directions in the Brillouin zone. (b) Fit of the conduction band energy calculated within the $G_0W_0$ approximation (red dots) using the Kane model (black line) along the $\Gamma$-X direction.}
\label{figureS_kane}
\end{figure}

Using the computational setup described below, we calculated a quasiparticle bandgap of 0.17\,eV, in agreement with prior $G_0W_0$ calculations starting from DFT/LDA (0.28\,eV~\cite{malone2013quasiparticle}), and HSE06 hybrid functional~\cite{heyd2003hybrid} (0.36\,eV~\cite{kim2009accurate}). The small difference between our calculation and prior works can be attributed to the dependence of $G_0W_0$ quasiparticle bandgaps on the mean-field starting point, as well as to different implementations of spin-orbit coupling effects. Fig.~\ref{figure:diff_delay}(c) depicts the quasiparticle band structure of InSb obtained from the interpolation of quasiparticle eigenvalues using Wannier functions~\cite{marzari1997maximally,souza2001maximally,yates2007spectral,mostofi2008wannier90}. As expected, we find that the conduction band is highly isotropic and can be described by the Kane model, for up to 400\,meV above the conduction band bottom, within numerical accuracy (see Fig.~\ref{figureS_kane}). Using this model, we calculated the effective mass at the conduction band bottom of 0.013\me, a value confirmed also by direct numerical calculations of the effective mass tensor, as described below, thereby further validating our computational approach and the use of the Kane model for our theoretical analysis. In addition, our calculated electron effective masses in good agreement with experiment (0.013\me~\cite{dresselhaus1955cyclotron}) and with prior first principles calculations (0.017\me~\cite{kim2009accurate}). 

\begin{figure}[ht!]
\includegraphics[width=0.45\textwidth]{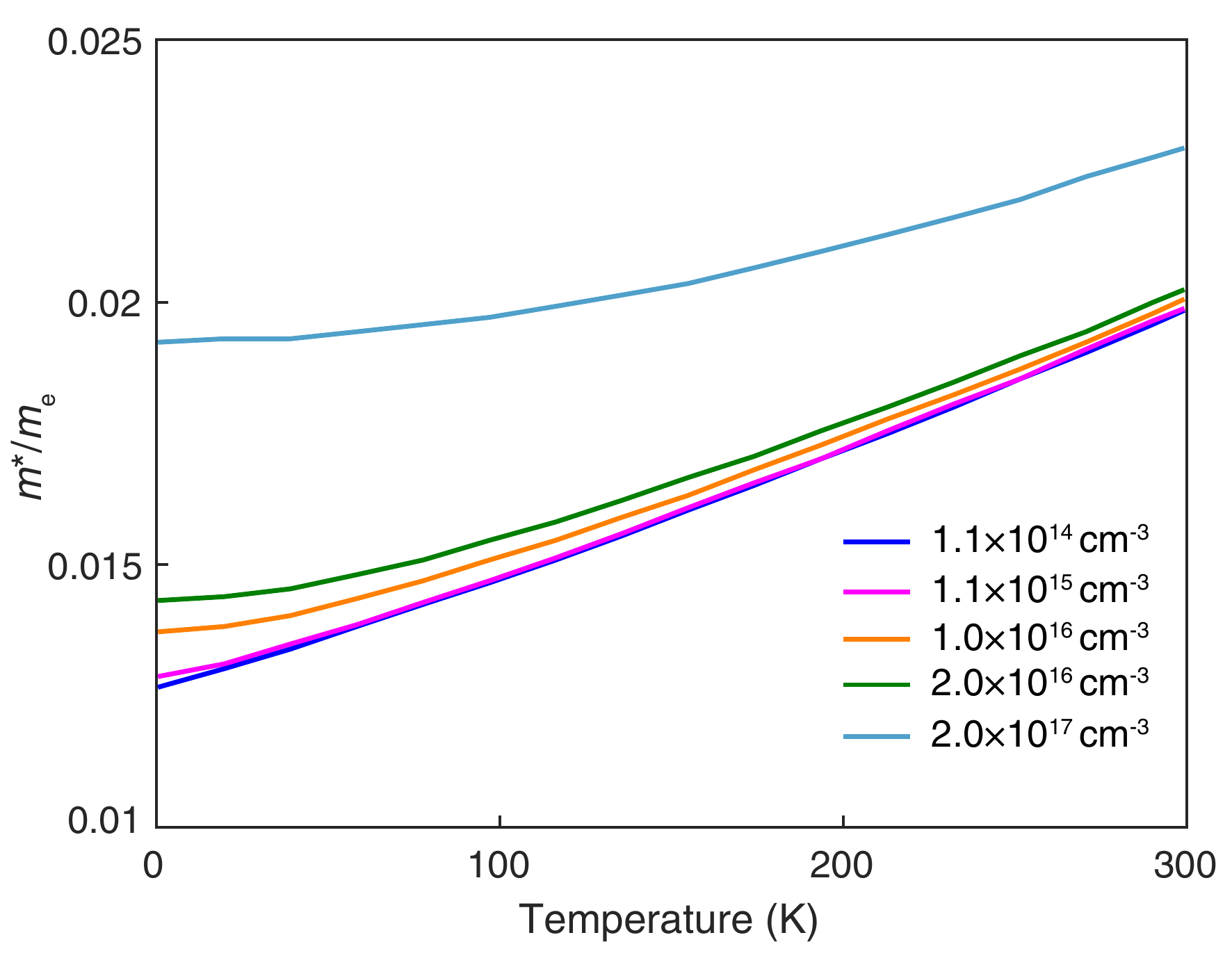}
\centering
\caption{Calculated effective masses of InSb as a function of temperature at different electron densities.}
\label{mass_temp}
\end{figure}

Having directly verified that the Kane model accurately describes the conduction band of InSb, we used the analytic expression of the conduction band dispersion to derive the density of states (including spin degeneracy), $g(E) = \frac{1}{2\pi^2} \Big(\frac{2m_0^*}{\hbar^2}\Big)^{3/2} [1+2E/E_{\rm g}]\sqrt{E(1+E/E_{\rm g})}$. The electron effective mass at various electron densities was then calculated as a function of temperature (Fig.~\ref{mass_temp}). Unlike what has been reported by Koteles and Datars~\cite{koteles1974temperature} where the effective mass was found to increase with temperature up to 55\,K and then decrease at higher temperatures, our calculations suggest a monotonic increase of effective mass with temperature. This is because in our study, the InSb sample was kept at a constantly low temperature (4\,K) and hence no lattice expansion effect was considered to change the band gap. In other words, the variable considered in our analysis is the electronic temperature which follows the Fermi-Dirac statistics, and results in the temperature-dependence of effective mass as shown in Fig.~\ref{mass_temp}.

\begin{figure*}[ht!]
\includegraphics[width=\textwidth]{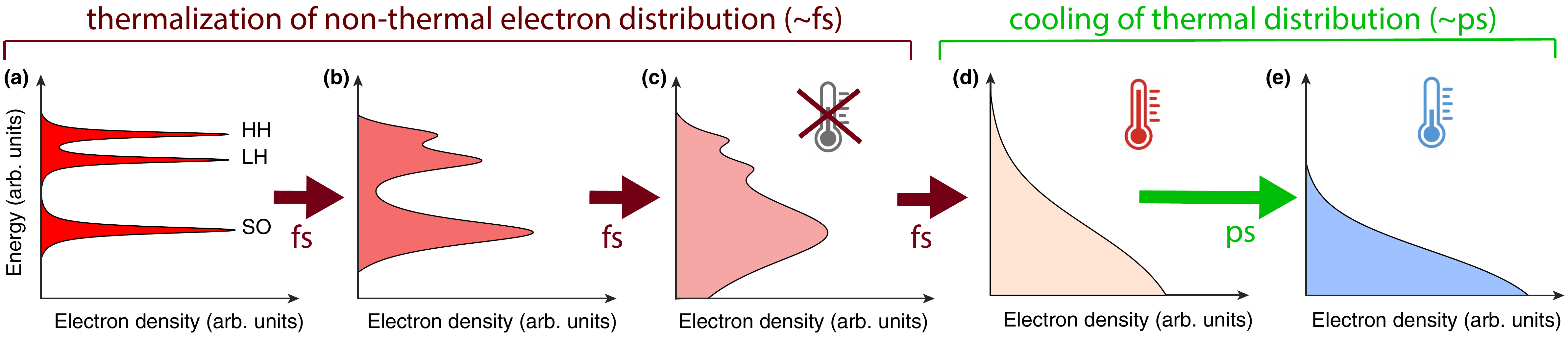}
\centering
\caption{Evolution of electron distribution in InSb conduction band before and after thermalization. Panels (a)--(c) show the non-thermal electron distribution with increasing time delays after photoexcitation on a femtosecond timescale. HH, LH and SO represent transitions from heavy-hole, light-hole and split-off bands respectively. Panels (d) and (e) show the thermalized electron distribution at high and low temperatures respectively, which follows the Fermi-Dirac distribution and happens on  a much slower timescale (picosecond) than the non-thermal distribution.}
\label{fig8}
\end{figure*}

It should be noted that although the hot electrons generated in the conduction band originate from heavy-hole (HH), light-hole (LH) and split-off (SO) bands as sketched in Fig.~\ref{figure:diff_delay}(d), our theoretical analysis focuses on the electron distribution after thermalization. As depicted in Fig.~\ref{fig8}(a), immediately after the 1.55\,eV femtosecond laser pulse is absorbed, the resulting conduction-band electrons have a peaked energy distribution owing to different electrons having been photoexcited from different bands  (HH, LH and SO bands). This energy distribution cannot be characterized by a single temperature and so is termed ``non-thermal''.  At later times, the peaked electron distribution start to broaden due to  thermalization (see Fig.~\ref{fig8}b and Fig.~\ref{fig8}c). This process happens on an ultrafast (femtosecond) timescale, during which the electron distribution is still non-thermal. In contrast, when the electrons have thermalized, they will follow the Fermi-Dirac distribution as shown in Fig.~\ref{fig8}(d). As the electron temperature decrease, the Fermi-Dirac distribution flattens, which is illustrated in Fig.~\ref{fig8}(e). Since our cyclotron resonance measurements were performed at 200\,ps after photoexcitation, the electrons in the conduction band of InSb would have thermalized and therefore, the different excitation efficiencies between HH, LH and SO bands were not relevant in our analysis. Instead, what we are interested here is the temperature-dependence of electron effective mass in InSb, which is illustrated in Fig.~\ref{mass_temp}. As shown in Fig.~\ref{figure:meff_temp}, our theoretical and experimental results show a good agreement with previous studies, which performed effective mass measurements on doped InSb samples.

\section*{Appendix F: Computational setup}
All calculations were performed using the experimental structure of InSb extracted from the Inorganic Crystal Structure Database (database code 162197; Ref.~\citenum{ersching2008structural}). We calculated the quasiparticle band structure of InSb within the $G_0W_0$ approximation~\cite{hybertsen1986electron}, including spin-orbit coupling, as implemented in the Yambo code~\cite{marini2009yambo}. We calculated the starting point mean-field eigenvalues within the generalized gradient approximation to density functional theory, in the Perdew-Burke-Erzerhof parametrization (DFT/PBE)~\cite{perdew1996generalized}, as implemented in the Quantum Espresso code~\cite{giannozzi2009quantum}. We used the ONCVPSP code to generate fully-relativistic optimized norm-conserving Vanderbilt pseudopotentials (ONCV)~\cite{hamann2013optimized}, using a similar setup to that reported in the Pseudo-Dojo database~\cite{van2018pseudodojo}. In our pseudopotentials we included semicore electrons for both In and Sb, in the following valence configuration: 4s$^2$4p$^6$4d$^{10}$5s$^2$5p$^1$ for In and 4s$^2$4p$^6$4d$^{10}$5s$^2$5p$^3$ for Sb. 

For the calculation of mean-field eigenvalues we used a plane wave cutoff of 100\,Ry and a $\Gamma$-centered $\mathbf{k}$-point grid of $20\times20\times20$. For the calculation of quasiparticle energies we used plane wave cutoffs of 66\,Ry and 20\,Ry for the exchange and polarizability respectively, 1500 bands and a $\mathbf{k}$-point grid of $10\times10\times10$ centered at $\Gamma$. This setup is similar to the computational setup reported in Ref.~\cite{malone2013quasiparticle}, and we estimate that the quasiparticle bandgap of InSb is converged within 30\,meV. 

 We calculated the quasiparticle band structure using Wannier interpolation, as implemented in the Wannier90 code~\cite{marzari1997maximally,souza2001maximally,yates2007spectral,mostofi2008wannier90}. We calculated the maximally localized Wannier functions by interpolating the single-particle DFT/PBE band structure using a coarse $\mathbf{k}$-point grid of $10\times10\times10$, starting from 16 $sp^3$ projectors. We then used the calculated Wannier functions to interpolate the quasiparticle band structure from quasiparticle eigenvalues calculated on the same mesh. We note that the four highest occupied quasiparticle eigenvalues at the $\Gamma$ point, which are expected to be degenerate in InSb, differ by up to 50\,meV, and we attribute this numerical inaccuracy to convergence of quasiparticle energies. To correct this behavior for the construction of the quasiparticle band structure, we set the top four quasiparticle eigenvalues at the top of the valence band equal to the average of the four calculated eigenvalues. This modification has no effect on the shape or position of the conduction band edge. 

We calculated the electron effective mass both from fitting the conduction band bottom to the Kane model, and through direct computation of the effective mass tensor, $m_{ij}^* = 1/\hbar^2 (\partial^2\epsilon/\partial k_i \partial k_j)^{-1}$, where $i$ and $j$ index the Cartesian directions. We calculated the second derivatives using the finite-difference method, with an increment of $\delta k = 0.001$\,$(2\pi/a)$, where $a$ is the lattice parameter of InSb. Both methods yield effective masses within $10^{-4}$\me.

\clearpage
\normalem
\end{document}